\documentclass[twocolumn,amsmath,aps,prd,preprintnumbers,amssymb,nofootinbib,superscriptaddress,showpacs,floatfix,balancelastpage]{revtex4-1}
\usepackage{amssymb}
\usepackage{amsmath}
\usepackage{graphicx,longtable,mathrsfs,color,array}
\usepackage{epsfig,subfigure,placeins,float} % plots
\usepackage{mathrsfs}
\usepackage{longtable}
\usepackage[usenames,dvipsnames]{xcolor}
\usepackage{mathtools}
\usepackage{relsize}
\usepackage{geometry}
\usepackage{graphicx}
\usepackage{hyperref}
\usepackage{color}
\usepackage{fullpage}

\begin{document}

\newcommand{\tkDM}[1]{\textcolor{red}{#1}}  % Doddy
\newcommand\T{\rule{0pt}{2.6ex}} 
\newcommand\B{\rule[-1.2ex]{0pt}{0pt}}
\newcommand{\be}{\begin{equation}}
\newcommand{\ee}{\end{equation}}
\newcommand{\apjl}{Astrophys. J. Lett.}
\newcommand{\aap}{Astron. Astrophys.}
\newcommand{\apjs}{Astrophys. J. Suppl. Ser.}
\newcommand{\sa}{Sov. Astron. Lett.}.
\newcommand{\jpb}{J. Phys. B.}
\newcommand{\natu}{Nature (London)}
\newcommand{\aaps}{Astron. Astrophys. Supp. Ser.}
\newcommand{\aj}{Astron. J.}
\newcommand{\aas}{Bull. Am. Astron. Soc.}
\newcommand{\mnras}{Mon. Not. R. Astron. Soc.}
\newcommand{\pasp}{Publ. Astron. Soc. Pac.}
\newcommand{\jcap}{JCAP.}
\newcommand{\jmat}{J. Math. Phys.}
\newcommand{\prep}{Phys. Rep.}
\newcommand{\jtep}{Sov. Phys. JETP.}
\newcommand{\plb}{Phys. Lett. B.}
\newcommand{\pla}{Phys. Lett. A.}
\newcommand{\jhep}{Journal of High Energy Physics}
\newcommand{\physrep}{Phys.~Rep.}   % Physics Reports

%\newcommand*{\rom}[1]{\expandafter\@slowromancap\romannumeral #1@}
%%%%% AUTHORS - PLACE YOUR OWN MACROS HERE %%%%%

%%%%%%%%%%%%%%%%%%%%%%%%%%%%%%%%%%%%%%%%%%%%%%%%

\title{Non-linear hydrodynamics of axion dark matter: relative velocity effects and ``quantum forces''}
\author{David J.~E.~Marsh}\email{dmarsh@perimeterinstitute.ca}
\affiliation{Perimeter Institute, 31 Caroline Street N,  Waterloo, ON, N2L 6B9, Canada}
\date{Draft version: \today}

\begin{abstract}

The non-linear hydrodynamic equations for axion/scalar field dark matter (DM) in the non-relativistic Madelung-Shcr\"{o}dinger form are derived in a simple manner, including the effects of universal expansion and Hubble drag. The hydrodynamic equations are used to investigate the relative velocity between axion DM and baryons, and the moving-background perturbation theory (MBPT) derived. Axions massive enough to be all of the DM do not affect the coherence length of the relative velocity, but the MBPT equations are modified by the inclusion of the axion effective sound speed. These MBPT equations are necessary for accurately modelling the effects of axion DM on the formation of the first cosmic structures, and suggest that the 21cm power spectrum could improve constraints on axion mass by up to four orders of magnitude with respect to the current best constraints.  A further application of these results uses the ``quantum force'' analogy to model scalar field gradient energy in a smoothed-particle hydrodynamics model of axion DM. Such a model can treat axion DM in the non-linear regime and could be incorporated into existing N-body codes.

\end{abstract}

\maketitle

\section{Introduction}

The particle nature of the dark matter (DM) is unknown, yet cosmological and astrophysical probes provide a wealth of information about the length and time scales over which it must form structure, providing bounds on the DM particle mass. In the case of thermal DM, structure formation excludes hot DM and limits warm (W)DM to have mass $m_W\gtrsim \mathcal{O}(\text{few keV})$, e.g. Refs.~\cite{2013PhRvD..88d3502V,2014MNRAS.442.1597S}. In the case of non-thermal DM, for example ultra-light axions (ULAs) or other scalar fields, e.g. Refs.~\cite{HuBarkana&Gruzinov2000,2000ApJ...534L.127P,2014ASSP...38..107S}, the limit on the mass is $m_a\gtrsim \mathcal{O}(\text{few})\times 10^{-23}\text{ eV}$, e.g. Refs.~\cite{2006PhLB..642..192A,bozek2014,Hlozek:2014lca}. Forthcoming astrophysical data, for example from the 21cm power spectrum, could extend these bounds substantially and possibly find evidence pointing to the particle nature of DM. Utilising this data requires models for DM on small scales, handling non-linear effects. WDM and self-interacting DM are fairly well studied in this regard, e.g. Refs.~\cite{bode2001,2014MNRAS.439..300L,2014MNRAS.444.3684V}, but much less is known about how to model ULAs and other scalar fields. The most advanced simulations in this field are those of Schive et al \cite{2014NatPh..10..496S} for studying dwarf galaxy sized objects and density cores. Alternative simulation models are necessary, both to confirm the accuracy of Ref.~\cite{2014NatPh..10..496S} simulations and to study new effects on different scales.

In this note I present a formalism, the non-linear hydrodynamic equations (NLHEs) in the non-relativistic limit. This is a useful model to compute the effects of ULAs on the 21cm power spectrum. The cut-off of power at small scales, $k\gtrsim \mathcal{O}(1)\text{ Mpc}^{-1}$, induced by ULAs satisfying $m_a>10^{-23}\text{ eV}$, via the effect on star formation and the baryon-DM relative velocity \cite{2010PhRvD..82h3520T,
2011MNRAS.418..906T,2010JCAP...11..007D,2012MNRAS.424.1335F,2014IJMPD..2330017F,2014MNRAS.438.2664S}, has a knock-on effect at large scales that could distinguish ULAs from CDM. The NLHEs also provide a possible method to model ULAs using smoothed-particle hydrodynamics (SPH), which is distinct from the method of Ref.~\cite{2014NatPh..10..496S} who use the Widrow-Kaiser \cite{Widrow&Kaiser1993} approach. The SPH method can be more easily incorporated into existing N-body codes, such as \textsc{gadget} \cite{2005MNRAS.364.1105S}, which already contain SPH modules to deal with baryon pressure. The method has been discussed recently in Ref.~\cite{2015arXiv150303869M}, and draws on ideas from Bohmian mechanics and quantum trajectories (e.g. Refs.~\cite{lopreore_wyatt,wyatt_trajectories}) to compute the anomalous pressure due to the scalar field gradient energy. I present the equations, and discuss various issues and links to other studies; possible algorithms are discussed in Refs.~\cite{wyatt_trajectories,2015arXiv150303869M}, and will be the subject of future discussion.

All linear power spectra are computed using \textsc{axionCAMB} \cite{Hlozek:2014lca}, a modified version of the publicly available \textsc{camb} code \cite{camb}. I work in units where $\hbar=c=1$.
\vspace{-0.1in}
\section{Hydrodynamic Equations}

The action governing an axion (or general scalar field)\footnote{I do not include self-interactions, as they are model-dependent. One can show that they are sub-dominant to gravity on linear and non-relativistic scales for an axion with the canonical cosine potential. Scalar field condensates with attractive and repulsive self interactions are discussed in e.g. Ref.~\cite{2011PhRvD..84f3518C}. Axions have an attractive quartic self-interaction, and thus no Thomas-Fermi limit. Other aspects of the self-interaction will be discussed in future work.} of mass $m_a$ in general relativity is
\be
S=\int d^4 x \sqrt{-g}\left[-\frac{1}{2}(\partial\phi)^2-\frac{1}{2}m_a^2\phi^2\right] \, .
\ee
Assuming the vanishing of anisotropic stress, the line element in Newtonian gauge in an expanding universe is
\be
ds^2=-(1+2V)dt^2+a^2(1-2V)d\vec{x}^2 \, ,
\ee
where $a$ is the cosmic scale factor and $V$ is the Newtonian potential. The Hubble rate $H=\dot{a}/a$. The scalar field $\phi$ obeys the Klein-Gordon (KG) equation
\be
(\Box-m_a^2)\phi=0 \, .
\ee
To first order in $V\sim \epsilon_{\rm NR}^2$ (weak-field limit, where $\epsilon_{\rm NR}$ is the perturbative parameter for relativistic effects) the D'Alembertian is
\be
\Box=-(1-2V)(\partial_t^2+3H\partial_t)+a^{-2}(1+2V)\nabla^2-4\dot{V}\partial_t \, .
\ee
In the limit that $H/m_a\sim \epsilon_{\rm WKB}$ we can use the WKB approximation to solve for $\phi$, giving
\be
\phi=(m_a\sqrt{2})^{-1} (\psi e^{-im_at}+\psi^*e^{im_at}) \, .
\ee
Taking WKB to first order implies $\dot{\psi}/m_a\psi\sim \epsilon_{\rm WKB}$. I also take the non-relativistic limit by letting the dispersion relation for wavenumber $k$ be $\omega=\sqrt{m_a^2+k^2}=m_a+km_a^2+\mathcal{O}(\epsilon_{\rm NR}^2)$, i.e. $k/m_a\sim \epsilon_{\rm NR}$ \citep{Seidel:1990jh}. Performing a double expansion to $\mathcal{O}(\epsilon_{\rm NR,WKB}^2)$, the non-relativistic limit of the KG equation gives the equation for the WKB amplitude $\psi$ (and also independently for $\psi^*$):
\be
i\dot{\psi}-3iH\psi/2+(2m_a a^2)^{-1}\nabla^2\psi-m_aV\psi=0 \, .
\label{eqn:schrodinger_hubble}
\ee

Eq.~(\ref{eqn:schrodinger_hubble}) is the Schr\"{o}dinger equation on an expanding spacetime. The Madelung form (see below) of this equation has been presented before in e.g. Refs.~\cite{2011PhRvD..84f3518C,2012A&A...537A.127C} by writing the standard Schro\"{o}dinger equation in the comoving frame. Expressing the equations in this frame is not the norm in studies of the Schr\"{o}dinger-Possion system \cite{Widrow&Kaiser1993,2014NatPh..10..496S,2014PhRvD..90b3517U}, where expansion effects are accounted for by rescaling the wavefunction by the background solution.

The axion energy density, $\rho_a$, is found from the energy momentum tensor
\be
T^\mu_{\,\,\,\,\nu}=g^{\mu\alpha}\partial_\alpha\phi\partial_\nu\phi-\frac{\delta^\mu_{\,\,\,\,\nu}}{2}(g^{\alpha\beta}\partial_\alpha\phi\partial_\beta\phi+m^2\phi^2) \, ,
\ee
with $T^0_{\,\,\,\,0}=-\rho_a$ giving 
\be
\rho_a = \frac{1}{2}[(1-2V)\dot{\phi}^2+m_a^2\phi^2+a^{-2}(1+2V)\partial^i\phi\partial_i\phi] \, .
\ee
to first order in $V$. Taking the same limits as before we find the leading order piece is 
\be
\rho_a=|\psi|^2+\mathcal{O}(\epsilon) \, .
\ee

Cosmological perturbation theory for scalar fields normally makes a background-fluctuation split at the level of the field, $\phi=\bar{\phi}+\delta\phi$ (e.g. Refs.~\cite{Hwang&Noh2009,2010PhRvD..82j3528M}). This, however, does not preserve the canonical form of the fluid equations at non-linear order in density fluctuations. Therefore, I make the background-fluctuation split at the level of the density, as is usual for CDM and baryons (e.g. Ref.~\cite{2002PhR...367....1B}). 

To do this, express the wavefunction, $\psi$, in polar co-ordinates as $\psi=Re^{iS}$ (known as Madelung form) and transform Eq.~(\ref{eqn:schrodinger_hubble}) into equations of motion for 
\be
\rho_a=R^2\, , \quad \vec{v}_a\equiv (m_a a)^{-1}\nabla S \, ,
\ee
where this defines the axion fluid velocity, $\vec{v}_a$. Separating real and imaginary parts of Eq.~(\ref{eqn:schrodinger_hubble}) gives
\begin{widetext}
\begin{align}
\dot{\rho}_a + 3H\rho_a+\frac{\nabla}{a}(\rho_a \vec{v}_a)&=0 \, , \label{eqn:unpert_axion_density}\\
\dot{\vec{v}}_a+H\vec{v}+\left( \vec{v}_a\cdot \frac{\nabla}{a}\right)\vec{v}_a&= -\frac{\nabla}{a}\left( V-\frac{1}{2m_a a^2}\frac{\nabla^2\sqrt{\rho_a}}{\sqrt{\rho_a}} \right)\, \label{eqn:unpert_axion_velocity}.
\end{align}
\end{widetext}
The background-fluctuation split can now be carried out as usual by writing $\rho_a=\bar{\rho}_a(1+\delta_a)$. This gives
\be
\dot{\bar{\rho}}_a+3H\bar{\rho}_a=0 \Rightarrow \bar{\rho}_a=\Omega_a\rho_{\rm crit}(a/a_0)^{-3} \, ,
\label{eqn:axion_background}
\ee
i.e. the axion background energy density behaves, in the WKB approximation, as a $w=0$ barotropic fluid (true for an oscillating field with a harmonic potential minimum \cite{Turner:1983he}). The fluctuations obey
\begin{align}
\dot{\delta}_a +a^{-1}\vec{v}_a\cdot\nabla\delta_a&=-a^{-1}(1+\delta_a)\nabla\cdot\vec{v}_a \, , \label{eqn:axion_conservation}\\
\dot{\vec{v}}_a+a^{-1}\left( \vec{v}_a\cdot \nabla\right)\vec{v}_a&= -a^{-1}\nabla (V+Q)-H\vec{v}\, \label{eqn:axion_euler} \, , \\
Q\equiv -\frac{1}{2m_a^2 a^2}&\frac{\nabla^2\sqrt{1+\delta_a}}{\sqrt{1+\delta_a}} \, ,\label{eqn:q_def_for_eoms}
\end{align}
where I have defined the ``quantum potential'' $Q$ that accounts for the scalar-field gradient energy. When $Q\rightarrow 0$, scalar fields have a Jeans instability \cite{1985MNRAS.215..575K}. 

A background-fluctuation split on the Einstein equations gives
\be
3H^2=8\pi G \sum_i\rho_i\, , \quad \nabla^2V=4\pi G a^2 \sum_i\rho_i\delta_i \, ,
\label{eqn:einstein}
\ee
where the second equation applies in the sub-horizon, $H/k\sim\epsilon_{\rm NR}$, limit (see, e.g., Ref.~\cite{1995ApJ...455....7M}), and the sums extend over all species, $i$.

Eqs.~(\ref{eqn:axion_background}--\ref{eqn:einstein}) are the complete NLHEs in the non-relativistic limit for axion/scalar-field DM. These equations differ from those for CDM only by the inclusion of the $Q$ term in the Euler equation, Eq.~(\ref{eqn:axion_euler}). They are valid in the non-relativistic limit even for non-linear density and axion field perturbations.\footnote{The limits taken are valid after axions begin oscillating ($H<m_a$), on sub-horizon but super-Compton length scales ($H<k<m_a$), and for weak-fields ($|V|<1$). The first two of these are generally valid during structure formation in the matter dominated era down to sub-kiloparsec scales for the allowed axion DM particle masses. The weak-field limit can be violated, of course, via the instability to black hole formation.} They thus illustrate the equivalence between axion DM and CDM on large scales (where $k/m_a\ll 1$), to all orders in Newtonian perturbation theory. They also provide the correct setting to apply all the tools of standard perturbation theory \cite{2002PhR...367....1B} and modern developments such as the effective field theory of large scale structures \cite{2012JHEP...09..082C} to axion DM.
\vspace{-0.1in}
\section{Applications}

\subsection{Relative velocity of axion DM and baryons}

The relative velocity between CDM and baryons can have an observable effect by suppressing star formation in the first structures at high-$z$ \cite{2010PhRvD..82h3520T,
2011MNRAS.418..906T,2010JCAP...11..007D,2012MNRAS.424.1335F,2014IJMPD..2330017F}. The inclusion of the quantum pressure, $Q$, in the Euler equation  for axion DM demands a separate treatment. The relative velocity between axion DM and baryons at recombination, $z_{\rm rec}\approx 1020$, can be computed in first order cosmological perturbation theory using a Boltzmann code. 

\begin{figure}[t]
\includegraphics[width=0.48\textwidth]{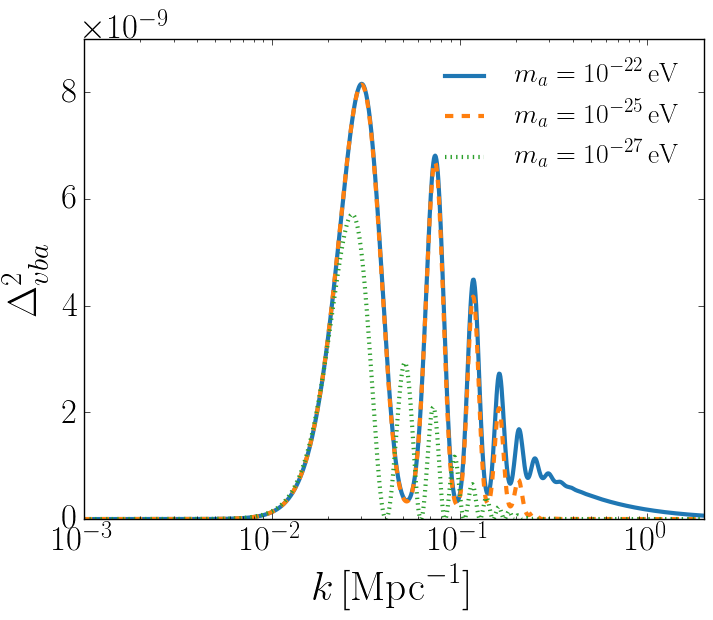}
\caption{Relative velocity power spectrum for axions and baryons (Eq.~\ref{eqn:vbc_power_def}) at $z_{\rm rec}=1020$ for various axion masses. On the scales shown the $m_a=10^{-22}\text{ eV}$ spectrum is indistinguishable from CDM.}\label{fig:vbc_rec_axion_masses}
\end{figure}

The variance of the relative velocity, $\vec{v}_{ba}=\vec{v}_b-\vec{v}_a$, is
\begin{align}
\langle v_{ba}^2\rangle&=\int \frac{dk}{k}\Delta^2_\zeta(k)\left(\frac{\theta_b-\theta_a}{k}\right)^2 \, ,\nonumber\\
&=\int \frac{dk}{k}\Delta^2_{vba}(k)\, ,\label{eqn:vbc_power_def}
\end{align}
which defines the relative velocity power spectrum, $\Delta^2_{vba}$, from the velocity divergence, $\theta_i\equiv a^{-1}\nabla\cdot\vec{v}_i$. The primordial curvature power spectrum is $\Delta_\zeta^2=A_s\left(\frac{k}{k_0}\right)^{n_s-1}$ with amplitude, $A_s$, and tilt, $n_s$, which are both well measured by \emph{Planck} \citep{planck_2015_params}; the pivot scale, $k_0$, is conventional. Fig.~\ref{fig:vbc_rec_axion_masses} shows $\Delta^2_{vba}$ for various axion masses. The lightest axions suppress the relative velocity power at large wavenumbers. On the scales shown, $m_a=10^{-22}\text{ eV}$ is indistinguishable from CDM.
\begin{figure}[t!]
\includegraphics[width=0.48\textwidth]{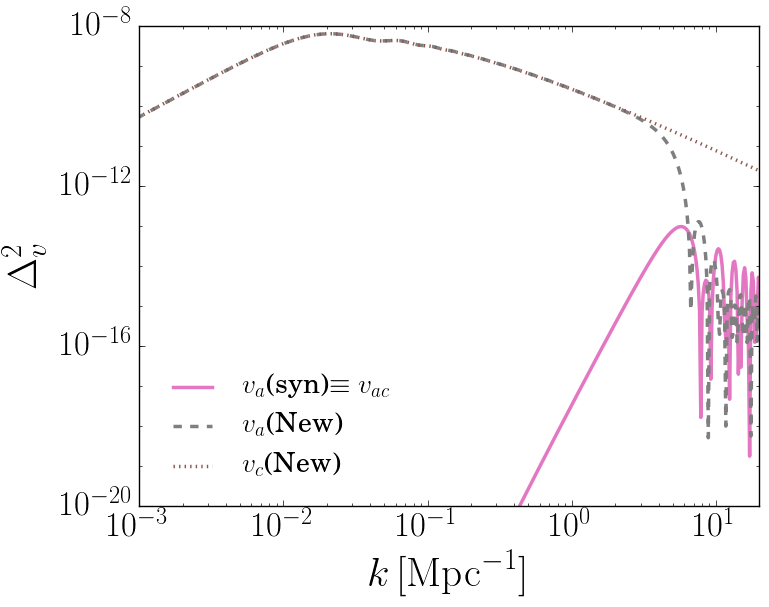}
\caption{Velocity power spectra for axions, $m_a=10^{-22}\text{ eV}$, and CDM in synchronous and conformal Newtonian gauges. Note that the synchronous gauge has $v_c(\text{syn})\equiv 0$. The gauge terms decay on sub-horizon scales.}\label{fig:vel_power_gauge}
\end{figure}

Relative velocities are gauge invariant, but the velocity of an individual species depends on the choice of gauge. For example, the CDM velocity divergence vanishes in synchronous gauge, $\theta_c(\text{syn.})=0$. Nevertheless it is instructive to look at the single-species velocity power. Fig.~\ref{fig:vel_power_gauge} shows the velocity power comparing CDM and an axion with $m_a=10^{-22}\text{ eV}$ in synchronous and Newtonian gauges \cite{1995ApJ...455....7M}. We observe that, in the Newtonian gauge axion DM has suppressed velocity power with respect to CDM for wavenumbers $k>k_{J,a}$, where $k_{J,a}=1.6 a \sqrt{Hm_a}$ is the axion Jeans scale. In the synchronous gauge, where $v_a$ is the relative velocity between CDM and axions, axions have velocity power for $k>k_{J,a}$. For $m_a<10^{-24}\text{ eV}$ the coherence scale of the axion-CDM relative velocity could be large and may serve as an additional probe of the composition of the DM, complementary to the CMB constraints of Ref.~\cite{Hlozek:2014lca} (see Refs.~\cite{2014arXiv1412.1660Z,2015arXiv150307480I} for the case of neutrino-CDM relative velocity).

The baryon-CDM relative velocity, $v_{bc}$, is coherent on scales $k_{\rm coh.}\gtrsim 1\text{ Mpc}^{-1}$. Large scale structure and the CMB \cite{bozek2014,Hlozek:2014lca} constrain ULAs to have $m_a>10^{-23}\text{ eV}$ if they are to be all of the DM. This implies that $k_{J,a}>k_{\rm coh.}$ and axion DM has the same large scale velocity relative to the baryons. Thus, moving background perturbation theory (MBPT), which treats the large scale relative velocity non-perturbatively, can be carried out for axion DM just as for CDM \cite{2010PhRvD..82h3520T} for cell sizes of order a few Mpcs. 

MBPT works by writing the velocities as
\be
\vec{v}_i(\vec{x},t)=\vec{v}_i^{\rm (bg)}(t)+\vec{u}_i(\vec{x},t) \, ,
\ee
then moving to the frame where
\be
\vec{v}_b^{\rm (bg)}=0 \text{ and }\vec{v}_a^{\rm (bg)}=-\vec{v}_{ba}^{\rm (bg)}\, .
\ee
and performing first order perturbation theory on the variables $\delta_i,\vec{u}_i$. Going to Fourier space and expanding to first order in $\delta_a$ the quantum pressure, $Q$, is
\be
Q\approx -\frac{k^2}{4m_a^2 a^2}\delta_a +\mathcal{O}(\delta_a^2) \, .
\ee
The equations of motion for the coupled axion-baryon system in MBPT are therefore
\begin{widetext}
\begin{align}
\dot{\delta}_a&=ia^{-1}(\vec{v}_{ba}^{\rm (bg)}(t)\cdot\vec{k})\delta_a-\theta_a \, , \label{eqn:dot_delta_a_vbc} \\
\dot{\theta}_a&=ia^{-1}(\vec{v}_{ba}^{\rm (bg)}(t)\cdot\vec{k})\theta_a-3H^2(\Omega_a(t)\delta_a+\Omega_b(t)\delta_b)/2 -2H\theta_a+k^4\delta_a/4m_a^2a^4\, ,  \\
\dot{\delta}_b&=-\theta_b \, ,  \\
\dot{\theta}_b&=-3H^2(\Omega_a(t)\delta_a+\Omega_b(t)\delta_b)/2-2H\theta_b +c_{s,b}^2k^2\delta_b/a^2 \, \label{eqn:dot_theta_b_vbc},
\end{align}
\end{widetext}
where $c_{s,b}$ is the baryon sound speed. In these equations we identify the axion effective sound speed:
\be
c_{s,a}^2\approx \frac{k^2}{4m_a^2 a^2}\, ,
\ee
which can be seen here as the manifestation of quantum pressure for linear density perturbations. 

The axion effective sound speed is the only term that distinguishes axion DM from CDM on small scales, suppressing power for wavenumbers $k>k_{J,a}$. Suppression relative to CDM is imprinted on the transfer function at matter-radiation equality, so the relevant scale is $k_{J,a}^{\rm eq}\approx 9 (m_a/10^{-22}\text{ eV})\text{ Mpc}^{-1}$ \cite{HuBarkana&Gruzinov2000}. The relative velocity of DM and baryons also suppresses power (in a given patch of size the $v_{bc}$ coherence length) over a range of wavenumbers centered near $k_{vbc}=aH/v_{bc}$ \cite{2010PhRvD..82h3520T}. The relevant ratio is thus $k_{J,a}^{{\rm eq}}/k_{vbc}$. At $z=40$ $k_{vbc}\sim 300\text{ Mpc}^{-1}$ and so
\be
(k_{J,a}^{{\rm eq}}/k_{vbc})|_{z=40}\approx 0.03 (m_a/10^{-22}\text{ eV})^{0.5} \, .
\ee

\begin{figure}[t]
\includegraphics[width=0.48\textwidth]{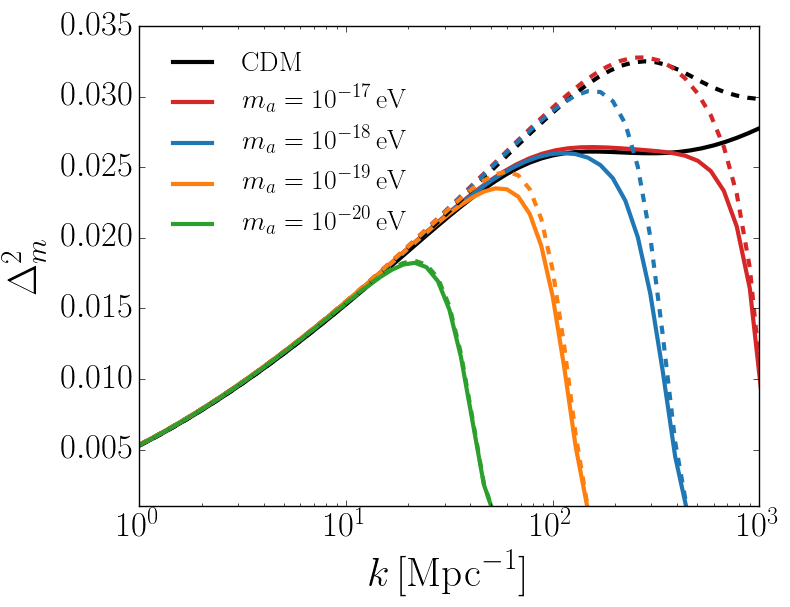}
\caption{Total matter power at $z=40$ with (solid lines) and without (dashed lines) the $v_{bc}$ effect, computed for various axion masses. For $m_a=10^{-18}\text{ eV}$ axion sound speed and $v_{bc}$ have effects on comparable scales.}\label{fig:dm2_vbc}
\end{figure}

I have verified these estimates-by-scale of the relevance of $k_{J,a}$ versus $k_{vbc}$ by numerical solution of Eqs.~(\ref{eqn:dot_delta_a_vbc}-\ref{eqn:dot_theta_b_vbc}), with the appropriate average over the $v_{ba}$ distribution, following Ref.~\cite{2010PhRvD..82h3520T}. The results are shown in Fig.~\ref{fig:dm2_vbc} at $z=40$ for various axion masses.\footnote{I use cosmological parameters of Ref.~\cite{planck_2015_params}.This accounts for the difference in normalization to Fig. 2 of Ref.~\cite{2010PhRvD..82h3520T}, who used a scale-invariant spectrum and different matter content.} For $m<10^{-18}\text{ eV}$ the axion sound speed cuts off power while the effects of $v_{ba}$ are only of order a few percent. For $m_a=10^{-18}\text{ eV}$ the axion sound speed cuts off power on scales where $v_{ba}$ is relevant at tens of percent in the power.

Planned 21cm power spectrum experiments may be able to detect the $v_{bc}$ effect for CDM \cite{2012Natur.487...70V,2013MNRAS.432.2909F,2014PhRvD..89h3506A}. The power spectrum shown in Fig.~\ref{fig:dm2_vbc} is averaged over the $v_{bc}$ distribution, but the power suppression caused by relative velocity is highly inhomogeneous. The relative velocity modulates the large-scale 21cm power spectrum by changing the small-scale matter power in regions of order the relative velocity coherence length: the moving background couples small and large scales. For example Ref.~\cite{2012Natur.487...70V} show that this modulation actually increases the 21cm power over a range of scales $k\sim 0.06\text{ Mpc}^{-1}$. The increase in power compared to the case where relative velocities are absent is larger than the projected sensitivity for telescopes such as the Low Frequency Array.

As shown in Fig.~\ref{fig:dm2_vbc}, axion DM with $m_a\lesssim 10^{-18}\text{ eV}$ wipes out small-scale power before modulation effects due to relative velocity become important. The small-scale power will be the same in different coherence length patches and there will be no additional modulation power communicated to large scales. For $m_a\lesssim 10^{-18}\text{ eV}$ the 21cm power on large scales will resemble the CDM power in the absence of relative velocity effects and the power boost shown in Ref.~\cite{2012Natur.487...70V} will be absent.  A detection of the modulation would exclude $m_a\lesssim 10^{-18}\text{ eV}$, an improvement of some four orders of magnitude with respect to current limits \cite{2006PhLB..642..192A,bozek2014}. This prognosis for 21cm agrees by order of magnitude with extrapolation from linear theory \cite{marsh2013b} using the WDM forecasts of Ref.~\cite{2014MNRAS.438.2664S} (although these forecasts ignored the $v_{bc}$ effect for WDM, and constraints were driven by the reduced collapse fraction, which is expected to be similar for axions and WDM \cite{marsh2013b}). On the other hand, the absence of a large-scale modulation will imply the existence of some small-scale cut-off in the power spectrum at $k<k_{vbc}$. 

While the small-scale power suppression caused by relative velocity is highly inhomogeneous, the power suppression caused by the axion sound-speed affects all of space. This could lead to further interesting effects. Finally, and very optimistically, the amplitude and shape of the large-scale 21cm power could probe axions in the range $10^{-18}\text{ eV}\lesssim m_a\lesssim 10^{-16}\text{ eV}$, where the small-scale power is affected on the same scales by both the axion Jeans scale and the $v_{bc}$ effect.
\vspace{-0.2in}
\subsection{Notes on an SPH treatment}

The identification of the scalar field gradient energy in the non-relativistic limit with a ``quantum force'' suggests a new approach to model axion/scalar-field DM on small scales. In quantum mechanics one can use the hydrodynamic Madelung representation to solve the Schr\"{o}dinger equation by introducing the quantum force. This is the so-called ``synthetic'' view of Bohmian mechanics \cite{wyatt_trajectories}: the motion of the hydrodynamic ``particles" is not necessarily understood as fundamental, and due to hidden variables, although it can nonetheless provide insights into processes such as quantum-mechanical tunneling \cite{lopreore_wyatt}.

In axion hydrodynamics, of course, the resemblance to Bohmian mechanics is just an analogy. Axions are described by a classical field due to the huge occupation numbers. One need not worry about measurement problems and decoherence. However, the hydrodynamic description is one of a condensate: although it is a classical field, the dynamics are ``very quantum'' from the particle viewpoint \cite{Guth:2014hsa}.

If cosmological structure formation is to be studied including the effects of the expansion and Hubble friction, Eqs.~(\ref{eqn:axion_conservation}-\ref{eqn:q_def_for_eoms}) should be used. If equations proposed by  Ref.~\cite{2015arXiv150303869M} are used, the expansion must be correctly accounted for by choice of frame. In the absence of the expansion-related terms, one can only study already-bound structures.  

It is conceptually simple to consider including the axion pressure in SPH. The quantum force in the hydrodynamic treatment is simply computed from the gradient of $Q$, and added to the force due to self gravity, giving the total force:
\be
F=-a^{-1}\nabla (V+Q) \, .
\ee
The gravitational potential, $V$, is found from the density by solving the Poisson equation as usual, while $Q$ is computed from gradients of the density via Eq.~(\ref{eqn:q_def_for_eoms}). Of course, the correct initial linear power spectrum for axion DM must be used. So far, so theoretically simple.

The numerical problem with such a treatment arises in two ways. Firstly, $Q$ depends on the Laplacian of the local density, and derivatives are difficult to estimate accurately in multi-scale problems like cosmological structure formation. The second difficulty arises because $Q$ can blow up in regions where $\rho_a\rightarrow 0$. This will be a problem in particular at interference nodes and in deep voids, where $\nabla^2\rho\neq 0$. In the quantum trajectories literature, many stable algorithms for computing the gradients have been studied, as well as methods for dealing with nodes \cite{wyatt_trajectories}, providing a possible ``off-the-shelf cookbook'' for cosmological simulators. .

As pointed out by Ref.~\cite{2015arXiv150303869M}, SPH is naturally adaptive, smearing the particles over large distances in low density regions and providing a natural force softening. In the scheme of Ref.~\cite{2015arXiv150303869M} the density field is represented with smoothing kernels at SPH particle locations:
\begin{align}
\rho_a(\vec{x})&=\sum_i \rho_a(\vec{x}_i) W(\vec{x}-\vec{x}_i;\xi) \, , \\
W(\vec{x};\xi)&=(\sqrt{2\pi \xi^2})^{-3}\exp (-|\vec{x}|^2/2\xi^2) \, , \\
\end{align}
where $\xi$ is a smoothing parameter. The advantage of this method is that density gradients only affect the kernel, whose derivative is analytically known.

In Ref.~\cite{2015arXiv150303869M}'s scheme, $\xi$ is chosen adaptively such that there is a fixed mass within the smoothing kernel. This ensures that regions of both high and low density are resolved equally well. However, since the quantum force can blow up inside low density regions such as voids, one might worry about the accurate resolution of the velocity field in these regions. Indeed, Ref.~\cite{2014NatPh..10..496S} showed that rapid oscillation of the wavefunciton is common everywhere, even in voids, implying large fluid velocity. The adaptive mesh-refinement in the solution of the Schr\"{o}dinger equation had to be carefully chosen to resolve phase oscillations. Similar considerations will apply to optimal choices of smoothing in SPH.

In the continuum limit, the SPH smoothing kernel is the position-space Husimi distribution (see, e.g., Ref.~\cite{ballentine_book})
\be
P_H(\vec{x})=\int d^3x' |\psi(x)|^2 W(\vec{x}-\vec{x}';\xi) \, .
\ee
The smoothing scale, $\xi$, is a free parameter reflecting the uncertainty relation. The Husimi distribution gives spatial resolution $\delta x=\xi$ and momentum resolution $\delta p=(2\xi)^{-1}$. The adaptive SPH smoothing described above will make $\xi$ large in regions of low density, leading to good velocity resolution in these regions and possibly mitigating any issues within voids. 

One sees that the choice of SPH smoothing is related to choosing the `best' measurements of a quantum probability distribution to extract phase-space information. Different possible choices of $\xi$ are suited to different problems. For example, $\xi$ could be chosen using the de Broglie scale, thus increasing resolution in both high density and high velocity regions. Such a smoothing might be useful for studying the phase-space density of axion DM halos \cite{2001ApJ...563..483T} and the structure of voids in this model. In quantum mechanics, the distribution is ``nearly classical'' if $\xi$ can be chosen such that $\delta x$ and $\delta p$ are each smaller than typical structures in the observables. Such a criterion could be used to determine when the quantum force is relevant, possibly saving computation time in an N-body simulation of this model. 

For further discussion of smoothing issues in the Schr\"{o}dinger picture, see Refs.~\cite{Widrow&Kaiser1993,2014PhRvD..90b3517U}; for smoothing issues in standard CDM, see e.g. Ref.~\cite{2015arXiv150305969M}. While smoothing is \emph{ad hoc} in CDM, it emerges naturally in the scalar field case, along with the adhesion approximation \cite{2011PhRvD..84f3518C}.

Since SPH modules are standard in many modern N-body codes (as they are used to model the baryon sound speed) I hope that this method for modelling axion DM on small scales can be implemented relatively easily into cosmological simulations. 
\vspace{-0.1in}
\section{Conclusions}
\vspace{-0.1in}
I have presented a simple derivation of the non-linear hydrodynamic equations (NLHEs), Eqs.~(\ref{eqn:axion_conservation}-\ref{eqn:q_def_for_eoms}) for axion/scalar field DM derived from a fundamental action. They are valid in the large occupation-number limit, where axion DM is described by the classical field equations. I have taken the non-relativistic, Newtonian limit, but have left the treatment of the axion energy density and field fluctuations non-perturbative. The equations presented differ from the usual Shcr\"{o}dinger picture for DM by the inclusion of expansion and Hubble drag effects explicitly, rather than through implicit frame choice. The NLHEs are the correct setting to study non-relativistic perturbation theory and non-linear clustering of axion DM. Two possible applications of these equations were outlined. 

The first discussed the relative velocity of baryons and DM. The effects shown in Fig.~\ref{fig:dm2_vbc} will delay formation of first stars and thus are expected to affect heating of the intergalactic medium and metal enrichment, and suggests that the 21cm power spectrum could tighten constraints on the axion mass by several orders of magnitude with respect to the current best constraints. Further study of this possibility is currently underway. If axions play a role in the formation of cores in dwarf galaxies, these searches will find evidence for axions \cite{2015arXiv150203456M}. Constraining $m_a\sim 10^{-18}\text{ eV}$ is a theoretically well motivated goal as this is the ``anthropic'' boundary for axions in string theory \cite{axiverse2009}, and is also a range of masses independently constrained by black hole spins \cite{Pani:2012vp,2015PhRvD..91h4011A}. 

The second application discussed an SPH model for N-body simulations of axion DM, which is an alternative to the adaptive-mesh-refinement approach of Ref.~\cite{2014NatPh..10..496S}, and could be incorporated within existing N-body codes. Work on this topic, too, is underway. Such simulations will open new doors on the study of structure formation with axion DM.
\\
\\
\indent\emph{Acknowledgments:} I acknowledge useful discussions with Simeon Bird, Brandon Bozek, Eddie Chua and Jonah Miller. I am especially indebted to Anastasia Fialkov, who made many helpful suggestions that inspired and improved this work, and to Yacine Ali-Haimoud and Daan Meerburg who provided code to explore the effect of $v_{bc}$ numerically. Research at Perimeter Institute is supported by the Government of Canada through Industry Canada and by the Province of Ontario through the Ministry of Research and Innovation.

\bibliography{axion_hydro}

\end{document}